# Inter comparison of the magneto transport of $La_{2/3}Ca_{1/3}MnO_3$: Ag/In polycrystalline composites

Rahul Tripathi[1], V.P.S. Awana[1,*], S. Balamurugan[2], R.K. Kotnala[1], RamKishore[1], H. Kishan[1] and E. Takayama-Muromachi[2]

[1]National Physical Laboratory, Dr. K.S. Krishnan Marg, New Delhi –110012, India

[2]Advanced Nano Materials Laboratory, National Institute for Materials Science (NIMS), 1-1 Namiki, Tsukuba, Ibaraki, 305-0044, Japan

**Abstract**

In this article, we report the synthesis, magneto transport features, and magnetization of polycrystalline $La_{2/3}Ca_{1/3}MnO_3:Ag_x/In_x$ composites with x = 0, 0.1, 0.2, 0.3 and 0.4. In case of Ag the temperature coefficient of resistance (*TCR*) near ferromagnetic (*FM*) transition enhances significantly with addition of Ag. The *FM* transition temperature ($T^{FM}$) is also increased slightly with Ag doping. Magneto-transport measurements revealed that magneto-resistance (*MR*) is found to be maximum near $T^{FM}$. Very sharp *TCR* is seen near $T^{FM}$ with highest value of up to 15 % for Ag (0.4) sample, which is an order of magnitude higher than as for pristine sample and is the best value yet reported for any polycrystalline *LCMO* compound. Increased *TCR*, $T^{FM}$ and significant above room temperature *MR* of $La_{2/3}Ca_{1/3}MnO_3:Ag_x$ composites is explained on the basis of improved grains size. Interestingly the $La_{2/3}Ca_{1/3}MnO_3:In_x$ composites behaved exactly in opposite way resulting in decreased $T^{FM}$, and *TCR* compared to pristine LCMO compound. In fact the grains morphology of $LCMO:In_x$ composites is seen inferior to pristine *LCMO* which is opposite to the $LCMO:Ag_x$ case.

\* e-mail: awana@mail.nplindia.ernet.in

*Keywords*: $La_{2/3}Ca_{1/3}MnO_3$: $In_x/Ag_x$ composites; Magneto-transport; and Magnetization

## I. INTRODUCTION

Lanthanum-calcium-manganite (LCMO) and its composites have been studied for their physical properties like colossal magneto resistance (CMR), *para*-ferro transition ($T^{FM}$) and metal-insulator (MI) behavior for the past few years.[1] Recently several research studies have been emphasized on the synthesis and development of high quality thin films for the infrared/bolometric detectors.[2-4] One of the most important desired characteristic of the infrared/bolometric detecting materials is the temperature coefficient of resistance (TCR) i.e. the variation of resistance with temperature. Not only the TCR must be high, but must exhibit a sharp transition over small temperature difference to act as switch on/off device for bolometric applications.[3-5]

In our recent studies we have reported high TCR in *LCMO*:Ag composites[6] with different doping levels. The presently reported work is basically an extension of our previous work by comparing the TCR of *LCMO*:Ag with *LCMO*:In (Indium) composites at different doping levels. This work is largely the inter comparison of *LCMO*:$Ag_x$ with *LCMO*:$In_x$ composites with different doping levels (x = 0-0.4), the synthesis part, micro-structure, magneto-transport, and the magnetization of $La_{2/3}Ca_{1/3}MnO_3$:$Ag_x$ and $In_x$ polycrystalline composites are discussed in detail. It is found that the $La_{2/3}Ca_{1/3}MnO_3$:$In_x$ composites behavior is exactly opposite to that of $La_{2/3}Ca_{1/3}MnO_3$:$Ag_x$, resulting in decreased $T^{FM}$, TCR and deteriorated grains morphology to pristine (x = 0) *LCMO* compound.

## II. EXPERIMENTAL

The polycrystalline $La_{2/3}Ca_{1/3}MnO_3$: $Ag_x$ and $In_x$ composites with x = 0, 0.1, 0.2, 0.3, 0.4 were synthesized through a solid-state reaction route from stoichiometric amounts of high purity (> 3 N) $La_2O_3$, $CaCO_3$, $MnO_2$, $Mn_2O_3$, metallic Ag, and $In_2O_3$. These raw materials were thoroughly mixed and ground in an agate mortar with pestle to obtain starting mixtures for the solid-state reaction. The mixed powders were calcined at 1000ºC, 1100ºC and 1200ºC in air for 24 hours each, subsequently pellets were pressed from the powders. Finally the pellets were annealed in air for 48 hours at 1300 ºC. Then the pellets were annealed in flow of oxygen at 800 ºC for 48 hours, and subsequently cooled slowly to room temperature over a span of 12 hours. The structure and phase purity of the samples were checked by powder X-ray diffraction with Cu $K_\alpha$ radiation. Magnetization measurements were carried out in a commercial magnetometer with the superconducting quantum interference device (MPMS-XL; Quantum Design) under applied field of 10 kOe. The transport measurements were carried out in a commercial Quantum Design physical property measurement system (QD-PPMS–6600) apparatus in the range of 5 to 400 K.

## III. RESULTS AND DISCUSSION

Figure 1 depicts the normalized resistance ($R_T/R_{400}$) versus temperature plots of *LCMO:In$_x$* composites. A decrease, though not monotonic ($T^{MI}In_{0.2} < T^{MI}In_{0.4}$) is seen in metal-insulator transition temperature ($T^{MI}$) with x in *LCMO:In$_x$* composites. Also the normalized resistance value near $T^{MI}$ in general increase with indium (In) doping, though again not monotonic, for example the $R_{T(peak)}/R_{400}$ value of $In_{0.20} > In_{0.40}$. The sharpness of the transition is also not improved with x in *LCMO:In$_x$* composites. *General trend of decreasing $T^{MI}$, sharpness of the transition and increase in $R_{T(peak)}/R_{400}$ near $T^{MI}$ with x for LCMO:In$_x$ composites, indicate* the possibility of indium being segregated or substituted at La-site instead of distributing at grain boundaries. Unlike in case of *LCMO:Ag$_x$* composites in *LCMO:In$_x$*, the Indium either gets substituted at La-site or remains in matrix as clusters. On the other hand in case of *LCMO:Ag$_x$* composites with x = 0, 0.1, 0.2, 0.3, 0., the peak normalized resistance decreases with x, though not monotonically, for example the $R_{T(peak)}/R_{400}$ value of Ag$_{0.20}$ > Ag$_{0.10}$. Interestingly, the $R_{T(peak)}/R_{400}$ trend followed at $T^{MI}$ is not the same at 5 K, which warrants further investigation. The sharpness of transition near metal-insulator- transition improves significantly with x. The sharpness of transition near metal-insulator transition temperature ($T^{MI}$) is generally defined by *TCR*, which we will discuss for all samples in next section. Further the $T^{MI}$ remains nearly invariant with doping percentage which is about 280 K for all x values in *LCMO:Ag$_x$* composites. This invariance in $T^{MI}$ can be explained on the basis that in our polycrystalline Ag samples, Ag has not substituted at La site. By substitution of Ag at La-site the $T^{MI}$ of pristine manganite compound increases to above room temperature. For example the $T^{MI}$ of La$_{1-x}$Ag$_x$MnO$_3$ and La$_{0.7}$Ca$_{0.3-x}$Ag$_x$MnO$_3$ is reported to increase with Ag substitution at La site[7,8].

The fine distribution of Ag at grain boundaries in *LCMO:Ag$_x$* composites results in tremendously improved grains morphology.[6] In this regards, the grains of *LCMO* pure and *LCMO:In$_x$* composites were compared and it was found that unlike as for *LCMO:Ag$_x$* composites,[6] the grain morphology of *LCMO:In$_x$* did not improve with indium (In) addition, please see Fig. 2 (a) for pristine and (b) for In$_{0.4}$ added *LCMO*. Note that the magnification is same for both figs. 2(a) and (b), but the morphology of pristine *LCMO* is far better than *LCMO:In$_{0.4}$*. This is just opposite to our earlier report[6] on *LCMO:Ag$_x$* composites. The morphology of pristine compound is seemingly same, *if compared on same scale*, as in ref.6 and current study, indicating similar quality of polycrystalline compounds in both cases.

As mentioned in the introduction, for infra-red detector/bolometric applications one requires the sharp metal-insulator-transition near room temperature[2,3]. The sharpness of the metal-insulator-transition near $T^{MI}$ is defined by temperature coefficient of resistance (*TCR*), where *TCR* = [1/*R*\*d*R*/d*T*]\*100. In *LCMO* composites high value of *TCR* is generally obtained near metal-insulator transition temperature. We determined the values of *TCR* for various composites from the *R*(*T*) data of *LCMO*:$Ag_x$ and *LCMO*:$In_x$ composites shown in Fig. 1 and its inset, respectively. Thus obtained *TCR* plots for typical samples, namely *LCMO*, *LCMO*:$Ag_{0.4}$, and *LCMO*:$In_{0.3}$ are shown in Fig. 3. For pristine *LCMO*, a *TCR* of around -2% is seen above its $T^{MI}$, i.e. in insulating regime. Below $T^{MI}$, i.e. in metallic regime, the *TCR* of pristine samples attains a maximum value of around 1%. The temperature difference between $TCR^{peak}$ values for pristine compound in insulating and metallic regimes is around 20 K. For *LCMO*:$In_{0.3}$ sample the $TCR^{peak}$ values are around –2.5% and 1% in insulating and metallic regimes, respectively. Interestingly, though the $TCR^{peak}$ values for *LCMO*:$In_{0.3}$ sample are comparable to the pristine sample, the temperature difference between two peak values is more than 50 K. This shows that as far as the infra-red detector/bolometric application is concerned, the *LCMO*:$In_{0.3}$ sample is much inferior in comparison to pristine *LCMO*. As far as the *LCMO*:$Ag_{0.4}$ is concerned its $TCR^{peak}$ values are around –1.5% and 15% in insulating and metallic regimes, respectively. Further, the temperature difference between these two peak values is around 5 K, which is one fourth to that of pristine and one tenth to that as for *LCMO*:$In_{0.3}$ samples. This suggests that *LCMO*:*Ag* proves to be a better option over *LCMO*:*In* composites as bolometric/infra-red detector material.

Magnetization versus temperature curves of *LCMO*:*Ag$_x$* and *LCMO*:*In$_x$* composites are shown in Fig.4 and its inset, respectively. All the samples exhibit usual paramagnetic to ferromagnetic transitions at $T^{FM}$ close to room temperature. For *LCMO*:*Ag$_x$* composites the $T^{FM}$ values remain about invariant at around 280 K with x. Also the saturation moment at 5 K is about 80 emu/g for the pristine and 120 emu/g for the Ag0.4 samples. On the contrary, in case of *LCMO*:*In$_x$* composites both the $T^{FM}$ (*with exception for $T^{MI}In_{0.2} < T^{MI}In_{0.4}$*) as well as the saturation moment at 5 K in *general* decreases with increase in x. The saturation moment of the sample at 5 K for *LCMO*:*In* is about 84 emu/g for pristine sample and decreases to 81, 76 and 69 emu/g for $In_{0.1}$, $In_{0.3}$ and $In_{0.4}$ samples, respectively. Increase in saturation moment at 5 K in *LCMO*:*Ag* composites suggests a the better coupling of ferromagnetic domains in this system in contrast to *LCMO*:*In* material where the same decreases with increase in x. To sum up one can conclude that *LCMO*:*Ag* composites are much superior to *LCMO*:*In*, with respect to their potential application to infra-red /bolometric detectors.

**ACKNOWLEDGEMENTS**

Authors from NPL appreciate the interest and advice of Prof. Vikram Kumar (Director) NPL in the present work. One of us Rahul Tripathi acknowledges the financial assistance provided by the UGC-India to carry out the research work.

**FIGURE CAPTIONS**

Figure 1. *R*(*T*) plots of *LCMO*:*Ag$_x$* composites, the inset shows the same for *LCMO*:*In$_x$ c*omposites.

Figure 2. (a) Scanning electron micrograph (*SEM*) of a *LCMO* pristine and *LCMO*:*In$_{0.4}$* samples at same magnification and scale.

Figure 3. Temperature coefficient of resistance (*TCR* %) for the selected *LCMO*:*Ag$_x$* and *LCMO*:*In$_x$* composite samples.

Figure 4. Magnetization versus Temperature (*M*/*T*) plots for the *LCMO*:*Ag$_x$* composites, the inset shows the same for *LCMO*:*In$_x$* composites

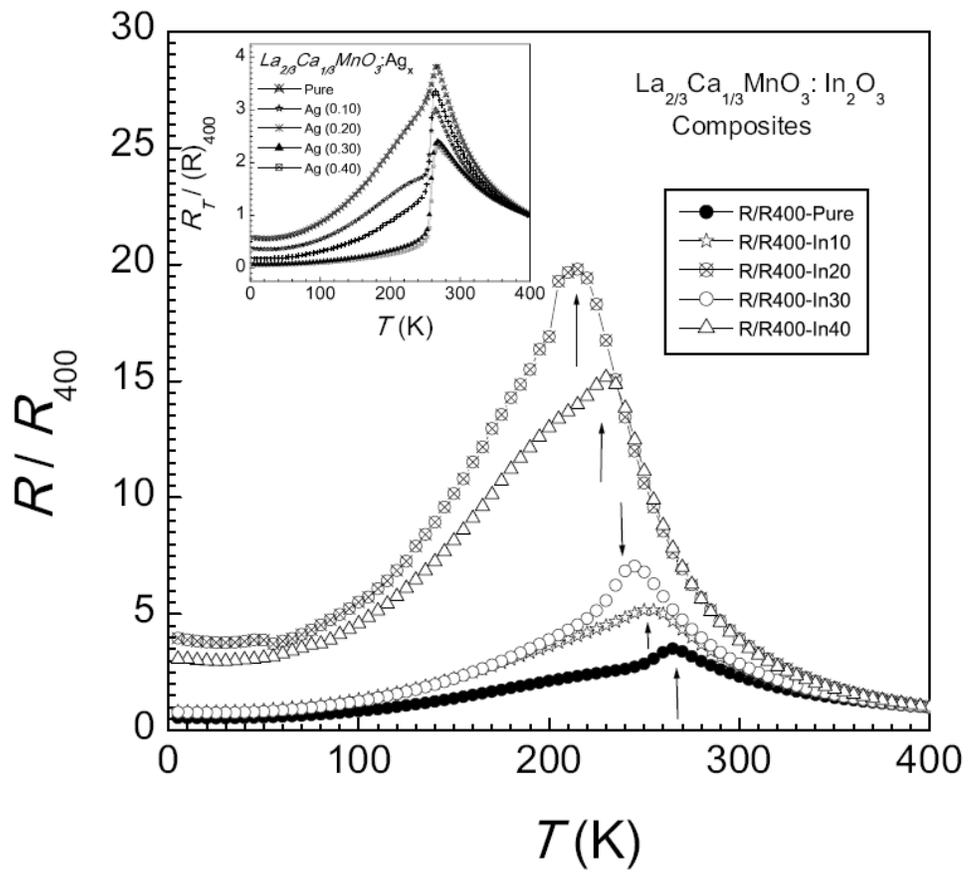

Fig.1. Tripathi *et al*.

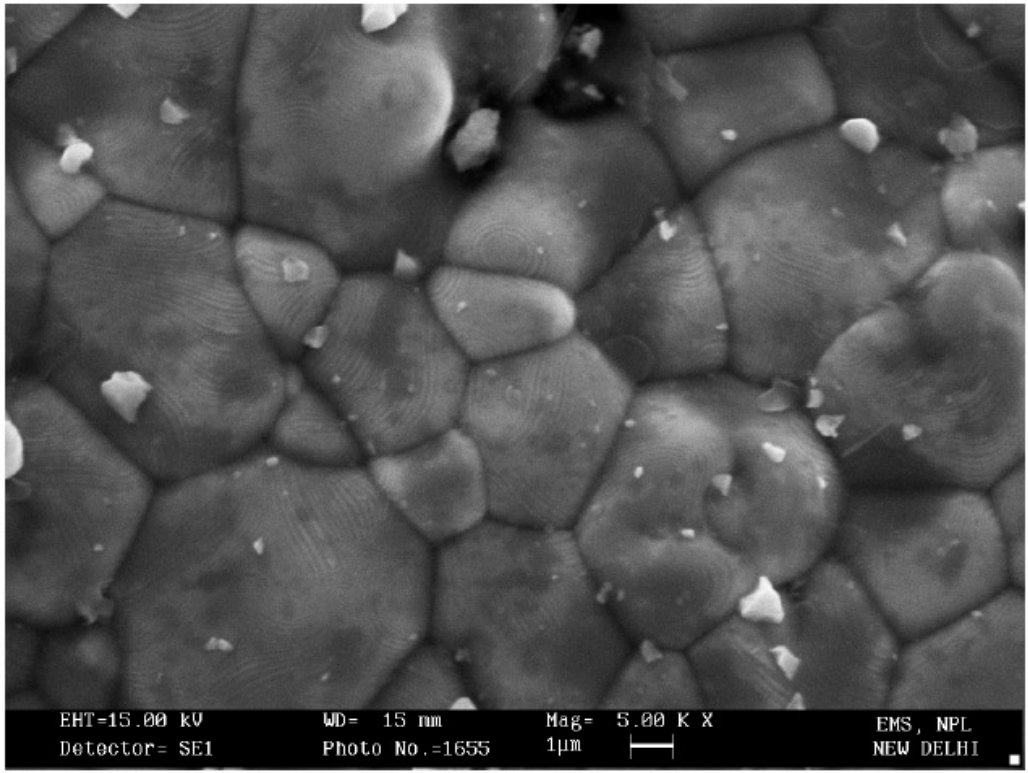

FIG. 2(a). Tripathi *et al.*

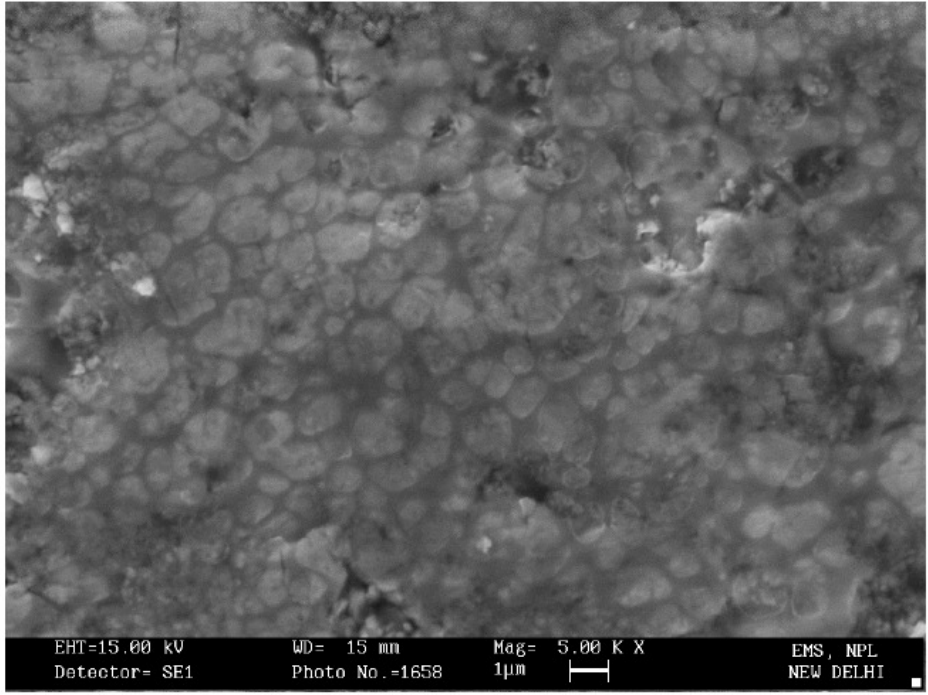

FIG. 2(b). Tripathi *et al.*

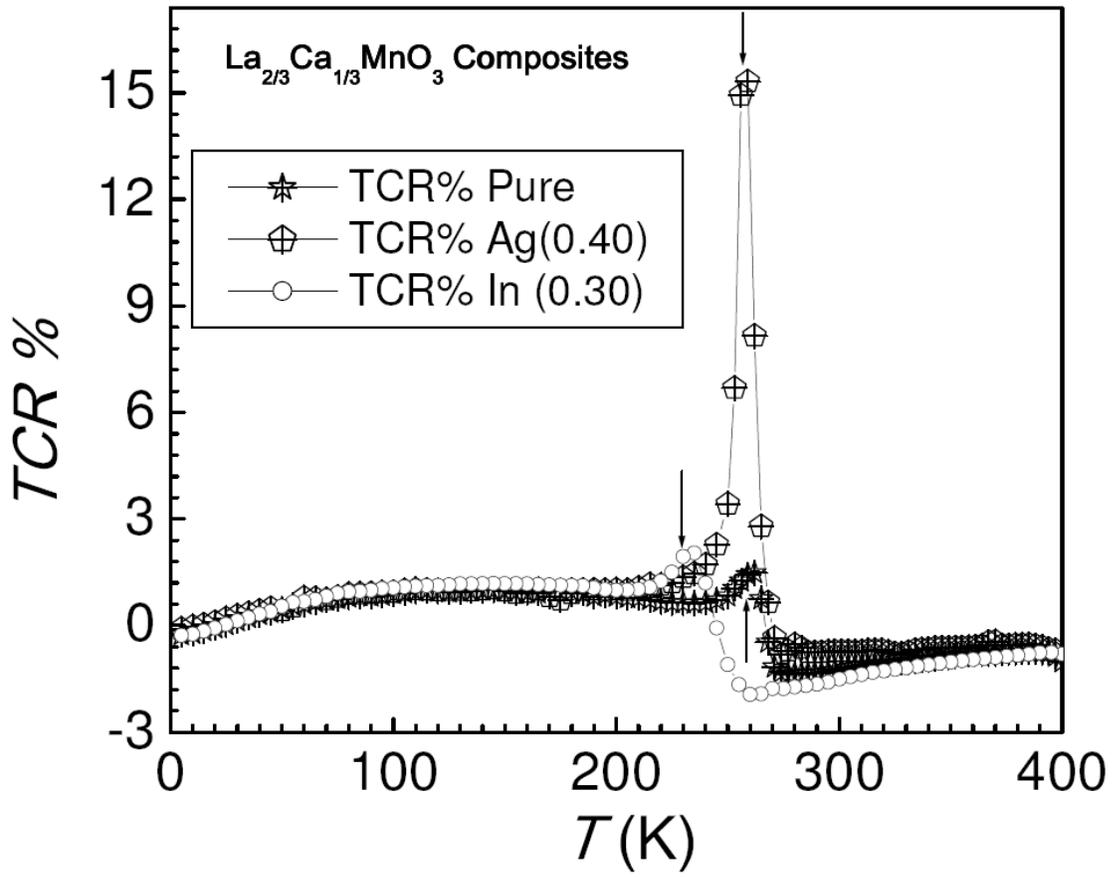

Fig.3. Tripathi *et al*.

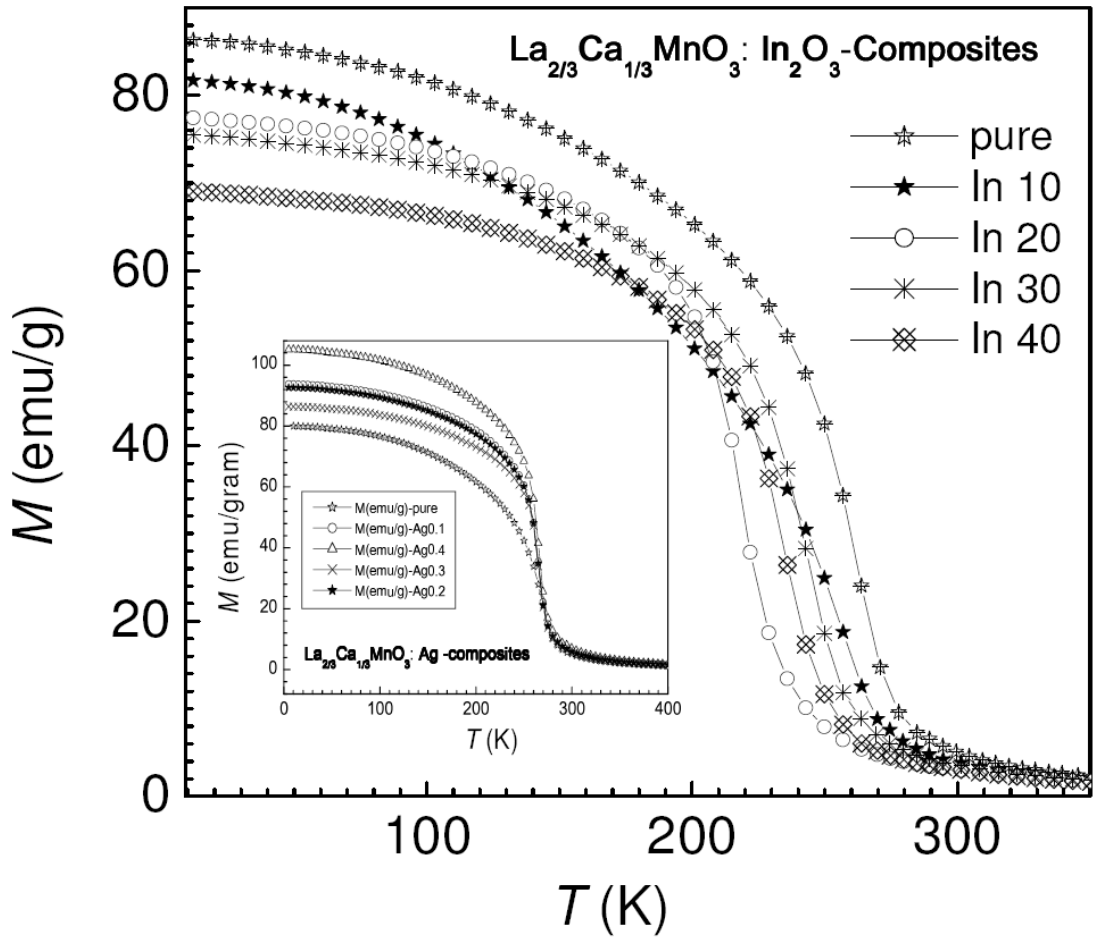

Fig.4. Tripathi *et al*.